%
%

\input harvmac.tex


%
\def\np#1#2#3{Nucl. Phys. {\bf B#1} (#2) #3}
\def\pl#1#2#3{Phys. Lett. {\bf #1B} (#2) #3}
\def\prl#1#2#3{Phys. Rev. Lett. {\bf #1} (#2) #3}
\def\physrev#1#2#3{Phys. Rev. {\bf D#1} (#2) #3}

\lref\krn{H. J. Kim, L. J. Romans and P. van Nieuwenhuizen, ``Mass Spectrum 
of Chiral Ten Dimensional $N=2$ Supergravity on $S^5$,'' 
\physrev{32}{1985}{389}.}%

\lref\gkp{S.S. Gubser, I.R. Klebanov, A.M. Polyakov,
``Gauge Theory Correlators from Non-Critical String Theory,''
hep-th/9802109.}%

\lref\witten{E. Witten, ``Anti-de Sitter Space and Holography,''
hep-th/9802150.}%

\lref\juan{J. M. Maldacena, ``The Large $N$ Limit of Superconformal Field
Theories and Supergravity,'' hep-th/9711200.}%

\lref\asty{O. Aharony, J. Sonnenschein, S. Yankielowicz and
S. Theisen, ``Field Theory Questions for String Theory Answers,''
hep-th/9611222, \np{493}{1997}{177}.}%

\lref\dls{M. R. Douglas, D. A. Lowe and J. H. Schwarz, ``Probing F
Theory with Multiple Branes,'' hep-th/9612062, \pl{394}{1997}{297}.}%

\lref\fs{A. Fayyazuddin and M. Spalinski, ``Large $N$ Superconformal
Gauge Theories and Supergravity Orientifolds,'' hep-th/9805096.}%

\lref\aoy{O. Aharony, Y. Oz and Z. Yin, ``M Theory on $AdS_p\times
S^{11-p}$ and Superconformal Field Theories,'' hep-th/9803051, to
appear in {\it Phys. Lett.} {\bf B}.}%

\lref\apsw{P. C. Argyres, M. R. Plesser, N. Seiberg and E. Witten,
``New $\cn=2$ Superconformal Field Theories in Four Dimensions,''
hep-th/9511154, \np{461}{1996}{71}.}%

\lref\intersect{M. Bershadsky and A. Johansen, ``Colliding
Singularities in F Theory and Phase Transitions,'' hep-th/9610111,
\np{489}{1997}{122}.}%

\lref\juansunny{N. Itzhaki, J. Maldacena, J. Sonnenschein and S.
Yankielowicz,
``Supergravity and the large N limit of theories with sixteen supercharges,''
 hep-th/9802042.}

\lref\senorientifold{ A. Sen, ``F-theory and orientifolds,'' 
Nucl. Phys. {\bf B475} (1996) 562, hep-th/9605150. }

\lref\krn{H. J. Kim, L. J. Romans and P. van Nieuwenhuizen, ``Mass Spectrum 
of Chiral Ten Dimensional $N=2$ Supergravity on $S^5$,'' 
\physrev{32}{1985}{389}.}%

\lref\gkp{S.S. Gubser, I.R. Klebanov, A.M. Polyakov,
``Gauge Theory Correlators from Non-Critical String Theory,''
hep-th/9802109.}%

\lref\witten{E. Witten, ``Anti-de Sitter Space and Holography,''
hep-th/9802150.}%

\lref\juan{J. M. Maldacena, ``The Large $N$ Limit of Superconformal Field
Theories and Supergravity,'' hep-th/9711200.}%

\lref\asty{O. Aharony, J. Sonnenschein, S. Yankielowicz and
S. Theisen, ``Field Theory Questions for String Theory Answers,''
hep-th/9611222, \np{493}{1997}{177}.}%

\lref\dls{M. R. Douglas, D. A. Lowe and J. H. Schwarz, ``Probing F
Theory with Multiple Branes,'' hep-th/9612062, \pl{394}{1997}{297}.}%

\lref\fs{A. Fayyazuddin and M. Spalinski, ``Large $N$ Superconformal
Gauge Theories and Supergravity Orientifolds,'' hep-th/9805096.}%

\lref\aoy{O. Aharony, Y. Oz and Z. Yin, ``M Theory on $AdS_p\times
S^{11-p}$ and Superconformal Field Theories,'' hep-th/9803051, to
appear in {\it Phys. Lett.} {\bf B}.}%

\lref\apsw{P. C. Argyres, M. R. Plesser, N. Seiberg and E. Witten,
``New $\cn=2$ Superconformal Field Theories in Four Dimensions,''
hep-th/9511154, \np{461}{1996}{71}.}%

\lref\ftheory{C. Vafa, ``Evidence for F Theory,'' hep-th/9602022,
\np{469}{1996}{403}.}%

\lref\morvaf{D. Morrison and C. Vafa, ``Compactifications of F Theory
on Calabi-Yau Threefolds 1,2,'' hep-th/9602114, hep-th/9603161,
\np{473}{1996}{74}, \np{476}{1996}{437}.}%

\lref\intersect{M. Bershadsky and A. Johansen, ``Colliding
Singularities in F Theory and Phase Transitions,'' hep-th/9610111,
\np{489}{1997}{122}.}%

\lref\seninter{A. Sen, ``A Nonperturbative Description of the
Gimon-Polchinski Orientifold,'' hep-th/9611186, \np{489}{1997}{139}.}%

\lref\blumzaf{J. D. Blum and A. Zaffaroni, ``An Orientifold from F
Theory,'' hep-th/9607019, \pl{387}{1996}{71}.}%

\lref\sw{N. Seiberg and E. Witten, ``Electric-Magnetic Duality,
Monopole Condensation and Confinement in $\cn=2$ Supersymmetric
Yang-Mills Theory,'' hep-th/9407087, \np{426}{1994}{19}.}%

\lref\swtwo{N. Seiberg and E. Witten, ``Monopoles, Duality and Chiral
Symmetry Breaking in $\cn=2$ Supersymmetric QCD,'' hep-th/9408099,
\np{431}{1994}{484}.}%

\lref\bds{T. Banks, M. R. Douglas and N. Seiberg, 
``Probing F Theory with Branes,'' hep-th/9605199, \pl{387}{1996}{278}.}%

\lref\argdoug{P. C. Argyres and M. R. Douglas, ``New Phenomena in
$SU(3)$ Supersymmetric Gauge Theory,'' hep-th/9505062,
\np{448}{1995}{93}.}%

\lref\gsw{M. Green, J. Schwarz and E. Witten, ``Superstring theory,''
Cambridge University Press (1987). }

\lref\gaunt{J.  Gauntlett, ``Intersecting branes,'' 
APCTP lectures, hep-th/9705011. }

\lref\ks{S. Kachru and E. Silverstein, ``4d Conformal Field Theories and
Strings on Orbifolds,'' hep-th/9802183, \prl{80}{1998}{4855}.}

\lref\gsvy{B. Greene, A. Shapere, C. Vafa and S.T. Yau, 
``Stringy cosmic strings and noncompact Calabi-Yau manifolds,''
Nucl. Phys. {\bf B337} (1990) 1.}

\lref\wilson{
J. Maldacena, ``Wilson loops in large N field theories,'' 
Phys. Rev. Lett. {\bf 80} (1998) 4859, hep-th/9803002;
 S-J.  Rey and  J. Yee,
``Macroscopic strings as heavy quarks in large N gauge theory and
anti-de-Sitter supergravity,''  hep-th/9803001. } 

\lref\bpsstates{A. Sen, ``BPS States on a Three Brane Probe,'' Phys. Rev. 
{\bf D55} (1997) 2501, hep-th/9608005; A. Johansen, ``A Comment on BPS States
in F-theory in 8 Dimensions,'' Phys. Lett. {\bf B395} (1997) 36,
hep-th/9608186; 
A. Fayyazuddin, ``Results in supersymmetric field theory from 3-brane
probe in F-theory,'' Nucl. Phys. {\bf B497} (1997) 101;
M. Gaberdiel and B. Zwiebach, ``Exceptional groups from open strings,''
Nucl.Phys. {\bf B518} (1998), hep-th/9709013;
I. Kishimoto, N. Sasakura ``M-theory description of BPS 
string in 7-brane background,'' hep-th/9712180;
O. Bergman and A. Fayyazuddin, ``String Junctions and BPS States in 
Seiberg-Witten Theory,'' hep-th/9802033;
Y. Imamura, ``$E_8$ flavor multiplets,'' hep-th/9802189;
A. Mikhailov, N. Nekrasov, and S. Sethi, ``Geometric Realizations of BPS
States in $\cn=2$ Theories,'' hep-th/9803142;
T. Hauer, ``Equivalent String Networks and Uniqueness of BPS States,''
hep-th/9805076;
O. De Wolfe, T. Hauer, A. Iqbal and B. Zwiebach, 
``Constraints on the BPS Spectrum of $\cn=2$ $d=4$ Theories with
ADE Flavor Symmetry,'' hep-th/9805220;
O. Bergman and A. Fayyazuddin, ``String Junction Transitions in the Moduli 
Space of $\cn=2$ SYM,'' hep-th/9806011.}

\lref\barton{M. R. Gaberdiel, T. Hauer, B. Zwiebach,
``Open string - string junction
transitions,'' hep-th/9801205.}

\lref\alex{ A. Kehagias, ``New type IIB vacua and their 
F-theory interpretation,'' hep-th/9805131.}

\lref\malstro{J. Maldacena and A. Strominger, ``$AdS_3$ Black Holes and
a Stringy Exclusion Principle,'' hep-th/9804085.}%

\lref\emil{ E. Martinec, ``Matrix Models of $AdS$ Gravity,''
hep-th/9804111.}

\lref\gubser{ S. Gubser, ``Can the Effective String see Higher Partial 
Waves?,'' Phys. Rev {\bf .D56} (1997) 4984,
hep-th/9704195.}

\lref\vafaetal{
M. Bershadsky and Z. Kakushadze, 
``String expansion as large N expansion of gauge theories,''
hep-th/9803076;
A. Lawrence, N. Nekrasov and C. Vafa, 
``On conformal field theories in four-dimensions,''
 hep-th/9803015;
Z. Kakushadze, ``Gauge theories from orientifolds and large N limit,''
 hep-th/9804184;
M. Bershadsky and A. Johansen, 
``Large N limit of orbifold field theories,''
hep-th/9803249.
} 

\lref\oztern{Y. Oz and J. Terning,
``Orbifolds of $AdS_5\times S^5$ and 4d Conformal Field Theories,''
hep-th/9803167.}

\lref\hortse{G. Horowitz and A. Tseytlin,
``On exact solutions and singularities in string theory,''
 Phys. Rev. {\bf D50} (1994) 5204, 
hep-th/9406067 and references therein.  
}

\lref\wittbaryon{
E. Witten, ``Baryons and Branes in anti-de Sitter Space,''
hep-th/9805112.
}

\lref\dasmu{K. Dasgupta and S. Mukhi, ``F-theory at constant coupling'',
 Phys. Lett. {\bf B385} (1996) 125,
hep-th/9606044.}

\def\cn{{\cal N}}
\def\IZ{\relax\ifmmode\hbox{Z\kern-.4em Z}\else{Z\kern-.4em Z}\fi}

\def\IR{\relax{\rm I\kern-.18em R}}
\def\tm{{\tilde m}}

%

%

\Title{\vbox{\baselineskip12pt
\hbox{hep-th/9806159}\hbox{HUTP-98/A046}\hbox{RU-98-26}}}
{\vbox{
{\centerline { The Large N Limit of 
${\cal N} =2,1 $ Field  Theories }}
\vskip .1in
{\centerline {from Threebranes in  F-theory   }}
  }}
\centerline{Ofer Aharony\foot{oferah@physics.rutgers.edu},
 Ansar Fayyazuddin\foot{ansar@schwinger.harvard.edu}
 and Juan Maldacena\foot{malda@bose.harvard.edu}}
\vskip.1in
\centerline{ $^1$ {\it Department of Physics, Rutgers University,
Piscataway, NJ 08855, USA}}
\centerline{$^{2,3}$ {\it Lyman Laboratory of Physics, Harvard University,
Cambridge, MA 02138, USA}}
\vskip.1in
\vskip .5in

\centerline{\bf Abstract}

We consider field theories arising from a large number of D3-branes
near singularities in F-theory. We study the theories at various
conformal points, and compute, using their conjectured string theory
duals, their large $N$ spectrum of chiral primary operators. This
includes, as expected, operators of fractional conformal dimensions
for the theory at Argyres-Douglas points. Additional operators, which
are charged under the (sometimes exceptional) global symmetries of
these theories, come from the 7-branes. In the case of a $D_4$
singularity we compare our results with field theory and find
agreement for large $N$. Finally, we consider deformations away from
the conformal points, which involve finding new supergravity solutions
for the geometry produced by the 3-branes in the 7-brane
background. We also discuss 3-branes in a general background.

\Date{}

%

\newsec{Introduction}

Through the conjecture \refs{\juan,\gkp,\witten} relating field
theories with string theory / M theory solutions, it has become
possible to explore the large $N$ limit of field theories by using
supergravity solutions.  One can motivate this relationship by
starting with a brane configuration in the full string theory and then
taking a low energy limit which decouples the field theory from
gravity, and at the same time considering the near horizon geometry of
the corresponding black brane supergravity solution.

In this paper we study ${\cal N}=2$ and ${\cal N} = 1$ field theories
which arise as worldvolume theories of D3-branes moving near
7-branes, i.e. 3-branes in F-theory \refs{\ftheory,\morvaf}.
In the simplest case F-theory reduces to a perturbative $\IZ_2$ orientifold
\senorientifold. The low-energy theory on the worldvolume of $N$
D3-branes is then \refs{\bds,\asty,\dls}
a $USp(2N)$ $\cn =2$ gauge theory with an
antisymmetric and four fundamental hypermultiplets. Deformations away
from the orientifold correspond to giving some masses to the
fundamentals.  When the masses are zero and the 3-branes are at the
origin (the orientifold point) we have a non-trivial conformal field
theory, whose large $N$ limit can be calculated by making an
orientifold of the $AdS_5 \times S^5$ gravity background. Note 
that this theory has dynamical quarks in the fundamental representation, 
so it can be used to study their behavior in the large $N$ limit. This system
was studied in \fs, together with other $\IZ_n$ orbifolds which do
not have a perturbative limit ($E_6,E_7,E_8$) that arise  as we bring
together many nonlocal 7-branes.  

In this paper we study these configurations directly, without using
their orientifold description.
In addition we can study
conformal field theories that occur  at Argyres-Douglas points
\refs{\argdoug,\apsw}. These cannot be viewed as orbifolds but
nevertheless we are able to find the large $N$ gravity description
for them. One of the novel features is that these theories have
operators with fractional conformal dimensions, which arise via some
simple generalization of spherical harmonics. One interesting implication
of these states is that we are forced to consider the
multiple cover of $AdS_5$. Though this is necessary in general to avoid 
closed timelike loops, the energies of states found in previous
computations, corresponding to operators
with  integer (or 1/2 integer) dimensions, were 
 consistent with periodic time.

In general, when the supergravity solution is singular, there will be
states coming both from the bulk fields and from the fields living on
the singularities. When the low-energy theory on the singularities is
weakly coupled, these states may be analyzed as easily as the bulk
modes, but this has not been done until now (except for a discussion
of tensor multiplet fields in \ks). In the theories we discuss there
will always be operators coming from states localized on the singular
surfaces, which in the orientifold case correspond to the fixed
surfaces. Topologically these singular surfaces look like an $S^3$
inside the $S^5$. We analyze the contributions to the spectrum from
the singularities as well as those of the bulk modes, checking them
against the expectations from the weakly coupled limit when it exists,
and finding agreement for large $N$.  Similarly, one can construct and
analyze configurations with two (or three) sets of intersecting
7-branes, which give rise to $\cn =1$ superconformal field theories
\refs{\seninter,\asty} about which not much is known.
Related configurations, that also describe conformal field theories,
 were studied in \alex .

Finally, we consider an arbitrary configuration of 7-branes.  It
has been shown in \refs{\bds,\asty,\dls} that (for theories with
$\cn=2$ SUSY) the metric seen by 3-brane probes moving in the
7-brane background is exactly the low-energy effective action for
the corresponding field theory.  We now go beyond the probe
approximation and consider the metric produced by the 3-branes. We
find that we can reduce the full non-linear gravity equations to a
single Laplace equation on the metric produced by the 7-branes.  We
discuss some qualitative features of the solution which do not require
solving the equation (a problem left for the future).  These solutions
provide examples of four dimensional non-conformal field theories
which can be solved in the large $N$ limit via supergravity.  In
particular one can have an asymptotically free theory with a
logarithmic running of the coupling.

In section 2 we review some aspects of the 7-brane geometry. We
begin by analyzing $\cn=2$ SCFTs which arise from D3-branes at 7-brane
singularities. In section 3 we calculate the operators coming from the
fields that live in the bulk of spacetime, while in section 4 we
calculate the contribution of the singular surfaces (the
7-branes). In section 5 we sketch the analysis for $\cn =1$
theories, and in section 6 we consider the supergravity solution in
the general non-conformal case.

\newsec{ D3-branes and Sevenbranes}

We consider the field theory arising when D3-branes move close to
7-branes. 
In the
field theory limit the 3-branes and 7-branes are at substringy
distances, so only the behavior near the singularities,
corresponding to the local F-theory geometry, is important. 
The simplest non-trivial case is when the 7-branes
correspond to a $\IZ_2$ orientifold point \senorientifold.  
A single
3-brane moving near a $\IZ_2$ orientifold point corresponds to a
$USp(2) \equiv SU(2)$ theory with four flavors \bds.  When we have $N$
branes we have \refs{\asty,\dls} a $USp(2N)$ theory with one
antisymmetric and four fundamental hypermultiplets.  If all the
7-branes and the 3-branes are  sitting together we have 
(at low energies) a
superconformal field theory. Since there is an $SO(8)$ gauge field
living on the 7-branes we have an $SO(8)$ global symmetry on
the D3-branes. This $\IZ_2$ singularity is called $D_4$.

We can get other theories by looking at different configurations of
the 7-branes. At generic points in 7-brane moduli space we will get
non-conformal field theories on the D3-branes. These theories will be
the subject of the last section. There are 7 types of singularities
which have a constant value of the dilaton and give rise to conformal
field theories on the D3-branes. They are the Argyres-Douglas points
$H_0,H_1$ and $H_2$ \refs{\argdoug,\apsw} which have $A_0,A_1$ and
$A_2$ gauge theories on the 7-branes (and corresponding global
symmetries on the D3-branes), and the $D_4, E_6, E_7$ and $E_8$
singularities, which give rise to corresponding gauge and global
symmetries \dasmu .
 Only the $D_4$ singularity can occur for any value of the
dilaton; the other singularities occur at a fixed value of
the string coupling, of order one.

The gravity description  D3-branes near 7-brane
singularities was derived in \fs\ for orbifold cases
($D_4,E_6,E_7,E_8$), and a similar description applies to all types of
singular 7-branes. For a D3-brane near a single 7-brane singularity,
the resulting metric is similar to the $AdS_5\times S^5$ metric, but
with the metric of the compact space, $d\tilde \Omega_5^2$,
 given by the angular variables of
\eqn\metricone{
ds^2 \equiv dr^2 + r^2 d\tilde \Omega_5^2 =
|dz|^2/|z|^\alpha + dx_3^2 +
dx_4^2 + dx_5^2 + dx_6^2
} 
instead of the $S^5$ metric, and with an
appropriate monodromy for the $B$-fields 
around the point $|z|=0$. The value of $\alpha$
depends on the singularity type; it is $\alpha={1\over 3},{1\over
2},{2\over 3},1,{4\over 3}, {3\over 2},{5\over 3}$ for
$G=H_0,H_1,H_2,D_4,E_6,E_7$ and $E_8$, respectively. Defining new
variables $w =
z^{1-\alpha/2}$ and $\tan^2(\theta) = |w|^2 / (x_3^2 + x_4^2 + x_5^2 +
x_6^2)$, we can write the angular part of \metricone\ as
\eqn\newmetric{ds^2 = d\theta^2 + \sin^2(\theta) d\phi^2 +
\cos^2(\theta) d\Omega_3^2 ~,} 
where $d\Omega_3^2$ is the metric on
$S^3$, $0 \leq \theta \leq \pi/2$ and $\phi=\arg(w)$ 
is periodic with period
$2\pi(1-\alpha/2)$, which corresponds to a full rotation in the $z$
plane. The $B$-fields generally have $SL(2,\IZ)$ monodromies
around this circle \senorientifold; for $D_4, E_6, E_7$ and $E_8$
these are described in \dasmu , and the $H_0,H_1$ and $H_2$ cases are
similar in this respect to $E_8, E_7$ and $E_6$, respectively. 
Except in the $D_4$ case, the string coupling $\tau$ also has monodromies
around this circle, which can be treated in a similar fashion.
The changed
periodicity and monodromies are the only difference between this solution 
and the $S^5$
solution, and it means that we have a $G$-type singularity at
$\sin(\theta)=0$, which is an $S^3$ in the compact space. This can be
interpreted as a $G$-type 7-brane sitting at $\sin(\theta)=0$ (and filling
$AdS_5$). The presence of the 7-branes breaks the $SO(6)$ isometry
symmetry (which is identified with the global symmetry in the four
dimensional field theory) to $SO(4)\times SO(2) \simeq SU(2)_R\times
SU(2)_L\times U(1)_R$.

The low-energy spectrum of the type IIB string theory on this space
will have a contribution from the bulk SUGRA modes on this space, and
a contribution from the singularities. 
We will calculate both of these contributions below.

Similarly, for the (orthogonal, SUSY-preserving) intersection of two
7-brane singularities \intersect, we will find the metric
corresponding to the angular variables of 
\eqn\metrictwo{ds^2 = |dz_1|^2 / |z_1|^{\alpha_1} + |dz_2|^2 / 
|z_2|^{\alpha_2} + dx_5^2 + dx_6^2,} 
where $\alpha_1,\alpha_2$ are the values of $\alpha$ corresponding to
each of the singularities. We will describe this in more detail below.
Notice that the values of $\alpha_i$ have to be compatible with one
another if both singularities exist only at fixed values of the
dilaton. In this case the isometry symmetry is broken to $U(1)^3$.

\newsec{Chiral Primaries of ${\cn =2}$ Superconformal Field Theories
 from Bulk Modes }

In this section we will compute the contribution to the low-energy 
spectrum on $AdS_5$ from
the supergravity modes, generalizing the analysis of \fs. 
As discussed above, these modes propagate on a
space which has the metric of an $S^5$ but with unusual boundary
conditions on one of the angular variables (denoted by $\phi$ in
\newmetric). Instead of being periodic with period $2\pi$, this
variable is periodic with period $2\pi(1-\alpha/2)$, with a monodromy
relating the fields at $\phi=0$ and at $\phi=2\pi(1-\alpha/2)$. As
discussed in \fs, the action of this monodromy on the supergravity fields can
be diagonalized, in which case it corresponds to a phase acting on
some of the fields. We will start by discussing scalar fields on the
compact space with no
non-trivial monodromy (for instance, the graviton fields $h_{\mu \nu}$
and $h^a_a$), and discuss the general case later.

The spectrum of the supergravity fields is determined by expanding the supergravity
equations (coupled to the 7-brane fields) to linearized order in the
fluctuations around the configuration described above. In the large
$g^2 N$ limit, the sphere is large, and we can ignore the interactions
of these fields with the 7-brane fields which live on a codimension
two subspace of the compact space. Since the metric on the
compact space is the same as the $S^5$ metric, we  find the same
equations as in the $S^5$ reduction of the supergravity fields \krn .
The only difference is the different periodicity of the
fields in the $\phi$ direction.

For massless 
scalar fields, the linearized equation of motion is simply the
Laplace equation on the compact space, whose solutions (for $S^5$) 
are the scalar spherical harmonics, in $k$'th symmetric traceless
representations of $SO(6)$ with eigenvalue $k(k+4)$. It is not hard to
derive this result also using the metric \newmetric\ (with standard
boundary conditions). The Laplacian in this metric is
\eqn\laplacian{
\nabla_5^2 = { 1\over \sin\theta\cos^3\theta} { d \over
d \theta} \sin\theta\cos^3\theta {d \over
d \theta }  - {m^2 \over \sin^2\theta } - { l(l+2) \over \cos^2 \theta }
}
where $m^2$  and $l(l+2)$ are  the eigenvalues of the Laplacian on $S^1$
and $S^3$ respectively. The latter has eigenfunctions
in the   $\bf{(l+1,l+1)}$ representations of $SU(2)_L\times
SU(2)_R$, with $l=0,1,2,\cdots$.
The $U(1)_R$ charge of the field is $2m$, and $m$ is an integer for
standard boundary conditions.
 Plugging in these eigenvalues we find a hypergeometric
equation in terms of  $\cos^2(\theta)$, which has a discrete series of
solutions that are regular at $\theta=0$ and at $\theta=\pi/2$,
labeled by $n=0,1,2,\cdots$. The total Laplacian eigenvalue
corresponding to the $n$'th solution is $k(k+4)$ where
$k=|m|+l+2n$. For a particular value of $k$ the states we find fill in
the $k$'th symmetric traceless multiplet of $SO(6)$. For instance, for
$k=2$ we find a $\bf{(1,1)}_{\pm 4}$ representation from $m=\pm 
2,l=0,n=0$, a
$\bf{(1,1)}_0$ representation from $m=0,l=0,n=1$, a $\bf{(2,2)}_{\pm 2}$
representation from $m=\pm 1,l=1,n=0$ and a $\bf{(3,3)}_0$ 
representation from
$m=0,l=2,n=0$. Together these form the $\bf{20^\prime}$ of
$SO(6)$.

How does this change when we implement the different periodicity
conditions described above ? The only thing that changes is that we
now have different possible values of $m$ from the $S^1$ Laplacian.
Instead of having $m$ be an integer from the eigenfunctions
$e^{im\phi}$, we now have $m=\tm/(1-\alpha/2)$ 
for integer $\tm$, corresponding to
the eigenfunctions $e^{i\tm \phi / (1 - \alpha/2)}$. Otherwise, the
analysis is the same as above. For the analysis
of the hypergeometric equation it does not matter if $m$ is an integer
or not. Thus, for any integer $\tm$ and non-negative integers $l$ and
$n$ we will find an eigenfunction of the Laplacian in the
${\bf(l+1,l+1)}_{2\tm/(1-\alpha/2)}$ representation, with an eigenvalue
$k(k+4)$, where 
\eqn\hyperg{k = {|\tm|\over(1-\alpha/2)} + l + 2n.}
The AdS/CFT correspondence relates a massless  
 field to an operator of scaling
dimension $\Delta=k+4$.
Some of the operators that were originally (in the $\cn=4$ theory)
in chiral primary multiplets
will remain in chiral primary multiplets also of the $\cn=2$
superconformal algebra, while others can be in large (non-chiral)
multiplets of the $\cn=2$ algebra. It is easy to identify which scalar 
fields
are lowest components of chiral primary multiplets of the $\cn=2$ algebra,
since these fields obey $\Delta=2j+R/2$ where $j$ is the spin of their
$SU(2)_R$ representation and $R$ is their $U(1)_R$ charge. Fields that
do not obey this relation can be either descendants of the chiral
primary fields or non-chiral fields.

Similarly, for scalar fields which have some phase from the monodromy,
again the only difference from the $S^5$ case will be the appearance
of different eigenvalues $m$ in the Laplacian, which will be shifted by
a constant compared to the values discussed above. As above, this will
change the $U(1)_R$-charge of the corresponding fields and their
dimension,
but otherwise the spectrum
remains unchanged. For instance, let us look at the fields arising
from a linear combination of $h^a_a$ and $D_{abcd}$, which in the
original $S^5$ case gave chiral primary fields of the $\cn=4$
superconformal algebra which were identified with $\tr(X^{i_1} X^{i_2}
\cdots X^{i_k})$ \foot{Note that the equation of motion
for this field includes also a linear (mass) term, causing the corresponding
dimensions to be $\Delta=k$ instead of $\Delta=k+4$ \refs{\krn,\witten}.}. 
Using the relation between the dimensions 
 the R-symmetry representations, it is easy to see that 
the condition for this  state to give a chiral primary field
of the $\cn=2$ superconformal algebra (namely, the field with
lowest dimension in an $\cn=2$ chiral primary multiplet) is that $n=0$. 
Consider the fields with $l=n=0$. In the
original $S^5$ case there was no mode with $k=0$ and the mode with
$k=1$ could be gauged away \krn\
so we only had $m = \pm 2, \pm 3, \cdots$. In the new metric it seems
that $\tm = \pm 1$ is also allowed (since it no longer corresponds to
a dimension one operator which would be a singleton field). 
Thus, we find a chiral primary field
in the $\bf{(1,1)}_{2\tm/(1-\alpha/2)}$ representation, with dimension
$\Delta = k = |\tm|/(1-\alpha/2)$, for any non-zero integer $\tm$. We can
identify these fields with the natural coordinates of the Coulomb
branch of these theories. For $N=1$, the field with $\tm=1$ can
be identified (in the usual sense of the AdS/CFT correspondence) with
the standard coordinate $u$ on the Coulomb branch, which was shown in
\apsw\ to indeed have dimension $\Delta = 1 / (1 - \alpha / 2)$
(namely, $\Delta=6/5$ for the $H_0$ case, $\Delta=4/3$ for $H_1$,
$\Delta=3/2$ for $H_2$, $\Delta=2$ for $D_4$, $\Delta=3$ for $E_6$,
$\Delta=4$ for $E_7$ and $\Delta=6$ for $E_8$). The fields with higher
values of $\tm$ may be identified with the natural coordinates
on the $N$'th symmetric product of the $u$ plane (for large
$N$), which is the Coulomb branch for the theory arising from $N$
D3-branes. Thus, our results agree with the field theory
expectations in this case. 
Similarly, we can analyze all the other supergravity fields,
and obtain predictions for the
full large $N$ spectrum of chiral primaries in all of these theories.

Note that our method of analyzing the spectrum is different from the
method of \fs, and apriori the results seem to be different. In
particular, in \fs\ the spectrum is described as a projection of the
original $S^5$ spectrum, while here we describe it as a shifted
version of the same spectrum. However, it is easy to check that in the
cases that correspond to orientifolds (namely, $\alpha=1,{4\over 3},
{3\over 2}$ and ${5\over 3}$) the
results actually agree. For instance, let us analyze the modes of a
periodic massless scalar field which correspond to operators of
dimension $\Delta=k+4$, for the case $\alpha=1$ 
which is a $\IZ_2$ projection. In
the analysis of \fs, we start with the original supergravity spectrum \krn\
in which we have such a mode in the $\bf{(l+1,l+1)}_{2m}$ representation
for $k=|m|+l+2n$, and then we project out the modes which have odd
values of $m$, so we find a field in the $\bf{(l+1,l+1)}_{4m}$
representation for $k=2|m|+l+2n$ for any integer $m$ and non-negative
integers $l$ and $n$. In our analysis, since in this case $m=2\tm$ for
integer $\tm$, we find exactly the same result by the method described
above. The same comparison works for all the other fields as
well. However, the analysis above applies also to cases which do not
have an orientifold description, such as the Argyres-Douglas points.

\newsec{Chiral Primaries of $\cn=2$ SCFTs from 7-brane Fields}

The compactifications of type IIB string theory which correspond to
$\cn=2$ SCFTs include, as described in \S2,
7-branes wrapped around an $S^3$ inside the
compact space (and filling the $AdS_5$ space). 
For the theory corresponding to
$N$ D3-branes at a $G$-type singularity, the low-energy theory on
these 7-branes is a 7+1 dimensional $\cn=1$ SYM theory of gauge group
$G$ (in order to get a conformal theory on the 3-brane the singularity can
be of type $G=H_0,H_1,H_2,D_4,E_6,E_7$ or $E_8$).
The low-energy spectrum of this compactification will thus
include the Kaluza-Klein modes of this vector multiplet on $S^3\times
AdS_5$, and by the usual AdS/CFT correspondence
\refs{\juan,\gkp,\witten} these will correspond
to primary operators in the SCFT. Since all these operators are in
small multiplets (with spins up to one), the corresponding operators
will necessarily be chiral primaries, and the AdS/CFT correspondence
implies that these will be all the chiral primaries charged under the
global symmetry
group $G$ which remain at finite dimension in the large $N$ limit (for
the $D_4$ theory we need to take also large $g^2 N$; for the other
theories $g_s$ is of order one so this is guaranteed).

In principle, we should compute the spectrum of masses of these states
by linearizing the equations of motion of the 7-brane fields in the
appropriate background, as was done for the supergravity fields in
\krn. However, since these states are all in small multiplets, their
masses (related to the dimensions of the corresponding chiral primary
operators in the SCFT) are completely determined in terms of their
R-symmetry representation. Thus, all we really need to do is compute 
which representations arise. 
The R-symmetry in the $\cn=2$ superconformal algebra is 
$SU(2)_R\times U(1)_R$, and a (scalar)
chiral primary field in an $SU(2)_R$ representation of spin $j$ 
and with $U(1)_R$ charge $R$
has dimension $\Delta = 2j + R/2$ (where $R$ is normalized so that
the SUSY generators have $R=1$). In the supergravity solution, this
R-symmetry is part of the isometry group of the compact space, which
is $SO(4)\times SO(2) \simeq SU(2)_R\times SU(2)_L\times U(1)_R$.

The 7+1 dimensional $\cn=1$ vector multiplet contains a vector field
$A_\mu$, a complex scalar field $z$ and fermions. We will concentrate
here only on the bosonic fields (obviously the fermionic spectrum is
related to this by the supersymmetry). The field $z$ is in the
$\bf{(1,1)}_2$ representation of the global symmetry group
$SU(2)_R\times SU(2)_L\times U(1)_R$. When we put this theory on
$S^3\times AdS_5$, the vector field will decompose into an
$AdS_5$-vector field $A_\mu$ and an $S^3$ vector $A_a$. The spectrum
of fields on $AdS_5$ will include the KK modes of all these fields.

For the scalar
field $z$ the expansion is $z(x,y) = \sum_k z_k(x) Y^k(y)$ where $x$
denotes the $AdS_5$ coordinates, $y$ denotes the $S^3$ coordinates,
and $Y^k(y)$ are the scalar spherical harmonics on $S^3$. On $S^d$,
these are in $SO(d+1)$ representations corresponding to symmetric
traceless products of ($\bf{d+1}$)'s. For $S^3$ we thus find that the
fields $z_k(x)$ are in the $\bf{(k,k)}_2$ representations of the
global symmetry group, for $k=1,2,3,\cdots$. 

For the vector field we have a similar expansion $A_\mu(x,y) = \sum_k
A_\mu^k(x) Y^k(y)$, leading to vector fields on the $AdS_5$ space in
the $\bf{(k,k)}_0$ representation for $k=1,2,3,\cdots$. 
For the internal components of the gauge field we have $A_a(x,y) =
\sum_k A_k(x) Y_a^k(y)$ where $Y_a^k$ are the vector spherical harmonics on
$S^3$. The fields $A_k(x)$ will thus be real scalar fields in the
$\bf{(k,k+2)}_0+\bf{(k+2,k)}_0$ representation of $SU(2)_R\times
SU(2)_L\times U(1)_R$ for $k=1,2,3,\cdots$. All these states are in the
adjoint representation of the gauge group $G$, and according to the
AdS/CFT correspondence they correspond to 
the only $G$-charged operators that remain at finite dimension in the
large $N$ limit.

Next, let us describe the supermultiplet structure that these fields
fall into. The supercharges of the $\cn=2$ theory are in the
$\bf{(2,1)}_1$ representation of $SU(2)_R\times SU(2)_L\times
U(1)_R$. There exist
small representations of the superconformal group that
start with a lowest component that is a real scalar in the
$\bf{k}_0$ representation of $SU(2)_R\times U(1)_R$. 
The other components of the multiplet arise by acting with supercharges $Q$
on this lowest component. We will describe here only the bosonic
components. Acting with two $Q$'s gives rise again to a scalar field,
in the $\bf{(k-2)}_{2}$ representation. The fact that we find a smaller
$R$-symmetry representation is a manifestation of the fact that this
is a chiral primary field in a small representation. Acting with two
${\bar Q}$'s gives the complex conjugate field. Acting with one $Q$
and one ${\bar Q}$ gives a vector field in the $\bf{(k-2)}_0$
representation. Finally, acting with two $Q$'s and two ${\bar Q}$'s
gives a scalar field in the $\bf{(k-4)}_0$ representation, and these are
all the bosonic fields in the short multiplet.

It is easy to see that all the fields found above fit into this type of
multiplet. Looking at singlets of $SU(2)_L$, we find that we have a
a real scalar field in the
$\bf{(3,1)}_0$, a complex scalar field in the $\bf{(1,1)}_2$
and a vector field in the $\bf{(1,1)}_0$, which fits into
the above structure with $k=3$
(in this case the highest
component of the multiplet vanishes). Similarly, the other fields fill
out other representations whose lowest component is a real scalar field in
the $\bf{(k+2,k)}_0$ representation, and which include a complex
$\bf{(k,k)}_2$ scalar
field, a vector $\bf{(k,k)}_0$ field, and a real $\bf{(k-2,k)}_0$ scalar
field, for $k=1,2,\cdots$. 
The dimension of the lowest component is determined by the
general formula for chiral primaries to be $\Delta = 2j+R/2 = k+1$,
and then the $Q^2$ components have dimension $k+2$ and the highest
component has dimension $k+3$.

In principle, we should compute the masses of all these fields by
expanding the 7-brane action, but this is not necessary since SUSY
guarantees that the masses will be the ones corresponding to these
dimensions. We should, however, comment on one apparent mystery, which
is that we find a real scalar field in the $\bf{(k+2,k)}_0$
representation with dimension $k+1$, while the $\bf{(k,k+2)}_0$ field
has dimension $k+5$. Naively, these fields arise from spherical
harmonics of the same Laplacian eigenvalue so they should have the
same mass, and therefore the operators should have the same scaling
dimension. However, while the corresponding vector spherical harmonics
indeed have the same eigenvalue of the Laplacian, they have
eigenvalues of opposite sign of the operator $*D$ defined by
$(*D Y)^a = \epsilon^{abc} D_b Y_c$,
where $a,b,c$ are $S^3$ coordinates and $D_b$ is the covariant
derivative. This operator squares to  the Laplacian, and the
eigenvalues of the
vector spherical harmonics in $\bf{(k,k+2)}$ representations have an
opposite sign from the eigenvalues of the $\bf{(k+2,k)}$ spherical
harmonics. This operator enters into the mass formula for the fields
through the coupling $\int d^8x F \wedge F \wedge D^{(4)}$ in the
7-brane action, where $F$ is the gauge field strength on the 7-brane
and $D^{(4)}$ is the 4-form RR field of type IIB string theory. This
leads to an equation of motion for the gauge field $A$ of the form
\eqn\ffd{\lform A^A \sim \epsilon^{ABCDEFGH} D_B A_C F^{(5)}_{DEFGH},}
where
$F^{(5)}$ is the self-dual 5-form field strength of $D^{(4)}$. In the
solution we are expanding around, $F^{(5)}$ is non-zero in the
$AdS_5$-components. Thus, we find a term of the form $\lform A^a \sim
\epsilon^{abc} D_b A_c$ in the equation of motion, which leads to a
shift in the mass of the field corresponding to the eigenvalue of $*D$,
in agreement with our results above.

Next, we would like to compare our results with the spectrum of
operators in the SCFT. The only case for which this is known is the
$D_4$ case, when the theory on the 3-branes is  an
$USp(2N)$ gauge theory with an anti-symmetric hypermultiplet and 4
fundamental hypermultiplets \refs{\bds, \asty,\dls}.
 The scalar fields in this theory include
a complex scalar field $X$ from the vector multiplet, which is in the
$\bf{(1,1,1,N(2N+1))}_2$ representation of the global and local 
symmetry group
\eqn\symmetry{SU(2)_R\times SU(2)_L\times SO(8)\times USp(2N)\times 
U(1)_R.}
The hypermultiplet scalar fields
are $q$ from the fundamental hypermultiplets, in the
$\bf{(2,1,8,2N)}_0$ representation, and $Y$ from the anti-symmetric
hypermultiplet, in the $\bf{(2,2,1,N(2N-1))}_0$ representation\foot{The 
$\bf{N(2N-1)}$ representation of $USp(2N)$ is actually 
reducible as $\bf{(N(2N-1)-1)}+\bf{1}$, with the {\bf 1} corresponding to the
center of mass motion of the 3-branes along the worldvolume of the
7-branes. Since this is a decoupled
free field we do not expect to see it in the bulk
gravity description.}. Both of
these fields obey reality conditions of the form $q^A_a =
\epsilon^{AB} J_{ab} (q^\dagger)_B^b$ and $Y^{AA'}_a = \epsilon^{AB}
\epsilon^{A'B'}
J_{ab} (Y^\dagger)_{BB'}^b$, where $A,B$ are $SU(2)_R$ indices, 
 $A',B'$ are $SU(2)_L$ indices, 
$a,b$ are
gauge group indices, and $J$ is the appropriate anti-symmetric
tensor. In the field theory the $SU(2)_L$ symmetry may be identified
with the flavor symmetry of the anti-symmetric hypermultiplet.

We are interested here only in operators which are charged under
$SO(8)$ (the other operators were described and compared with the AdS
construction in \fs, as described in the previous section). 
Obviously, these operators must involve at least
two squark fields $q$, or the fermions $\psi_q$ in the quark multiplets.  It
is easy to see that any operator with more than two quarks will be the
product (or the sum of products) of more than one gauge invariant
field, so it should not be compared with single-particle states in the
AdS theory. Thus, it is enough to look at operators with two quark fields.
The simplest such operator is just a product of two $q$'s.
Obviously, the product must be anti-symmetric in the $USp(2N)$ indices
to give a gauge-invariant field, so it can either by symmetric in the
$SO(8)$ indices and anti-symmetric in the $SU(2)_R$ indices or vice
versa. In the first case we get a field in the $\bf{(1,1,35+1)}_0$ of the
global symmetry, but this is not a chiral primary, since it does not 
have any R-charge\foot{Therefore, 
it does not obey the relation between the dimension
and R-charge for a chiral operator in the free theory. Alternatively we
can see that by acting on
it with SUSY generators we get a field in the (larger) $\bf{(2,1,35+1)}$
representation.}. 
On the other hand, in the second case we get a real
scalar field in the $\bf{(3,1,28)}_0$ representation, and we can
identify it with the ($\Delta=2$) field of the same representation 
that we found above. The rest of this representation can now be
constructed by acting on this operator with the supercharges. The
$Q\bar{Q}$ component will be exactly the global $SO(8)$ current (with
$\Delta=3$), while the
$Q^2$ component involves terms of the form $qXq$ and $\psi_q \psi_q$.

The other fields described above can similarly be constructed from the
product of $k-1$ $Y$-fields with the two quarks, starting from $q^A_a
Y^{ab}_B q^C_b$ where the $SU(2)_R$ indices $A,B,C$ must be multiplied
symmetrically to get a chiral field, and the $SU(2)_L\times SO(8)$
indices (in the $\bf(2,28)$ for $k=2$) were suppressed. It is slightly less
trivial to show that these are the only chiral primary fields in the
theory. Any product of $X$ with $q$ turns out to be a descendant
because of the $W = q X q$ superpotential. Similarly, any
anti-symmetric combination of $Y$'s is a descendant because of the $W
= Y X Y$ piece of the superpotential (the equation of motion of $X$
enables us to replace it by two $q$'s, allowing us to decompose the
field into a product of more than one gauge-invariant field). Thus,
for the $D_4$ case we find an exact agreement between the AdS
prediction for the spectrum and the field theory results for large
$N$, which can be viewed as evidence for the conjecture of \juan.
In both cases, the spectrum of ($G$-charged) chiral primary
fields includes the global symmetry current multiplet, and an infinite
(for large $N$) series of copies of this multiplet, with increasing
dimensions and $SU(2)_R\times SU(2)_L$ representations.

For finite $N$, $k$ cannot be arbitrarily large in the field
theory, since for $k$ of order $N$ the products of $k$ fields are no
longer independent (in particular, for $N=1$ $Y$ is a singlet, and
only the $k=1$ fields are independent). However, as discussed in \aoy,
it is not clear how to see this in the AdS construction. Similar bounds
were discussed for $AdS_3\times S^3$ in 
 \refs{\malstro,\emil,\gubser}. For other
groups $G$, the construction gives us a prediction for the spectrum of
chiral primaries which it is not clear how to check directly in field
theory. For $G=H_0,H_1,H_2$ we can construct the corresponding theory
by flows from the $D_4$ theory \apsw, which enables us to construct the
chiral primaries as subsets of the primaries of the $D_4$ theory. For
$G=E_6,E_7,E_8$ it is not known how to construct these spectra directly
from field theory; perhaps the predictions of the AdS/CFT
correspondence may be tested in string theory. 
We find the same
structure for the spectrum of $G$-charged states in all of these cases, 
with the only difference being in the global symmetry representation.

\newsec{${\cal N} =1$ Superconformal Field Theories
 from Intersecting Sevenbranes}

In this section we study the conformal field theories arising when
3-branes sit at the singularity in \metrictwo. We will describe the
general procedure for computing the large $N$ spectrum of these SCFTs,
but we will not compute them explicitly. The results here may easily
be generalized also to the case of three intersecting 7-branes.

\subsec{Bulk Contribution}

The analysis is similar to what we did above for
${\cal N} =2$ theories. 
We can choose coordinates so that the angular part of \metrictwo\ becomes
\eqn\intermetric{ds^2 = d\theta^2 + \sin^2(\theta) d\psi^2
+ \cos^2(\theta) d\phi_1^2 + \sin^2(\theta)
\cos^2(\psi) d\phi_2^2 + \sin^2(\theta) \sin^2(\psi)
d\phi_3^2,}
with $\alpha$-deformed boundary conditions and monodromies
in the $\phi_1$ and $\phi_2$
variables. The spherical harmonics will now be labeled by the momenta
$m_1,m_2$ and $m_3$ in the $\phi$ variables, and by two additional
non-negative 
integers $m$ and $n$, such that the eigenvalue of the Laplacian is
$k(k+4)$ where $k = |m_1| + |m_2| + |m_3| + 2m + 2n$. 
In the $S^5$ case all these numbers are integers. 
In our case the periodicity conditions on $\phi_1,\phi_2$ are different,
so $m_i = \tm_i/(1-\alpha_i/2)$ ($i =1,2$), but $m_3,n,m$ are still integers.
The $U(1)_R$ charge in the $\cn=1$ superconformal algebra 
is the sum of the $SO(2)$'s 
acting on $\phi_1,\phi_2,\phi_3$,
so it will be $2 m_1 + 2 m_2 + 2 m_3$.

Of course, all these shifts affect only fields which are charged under
the $SO(2)$ symmetries in the spaces transverse to the singularities. In
particular, the fields which are uncharged under the R-symmetry, such as
the energy-momentum tensor (the zero mode of  the graviton) and the
global symmetry currents (the zero modes of the 7-brane gauge fields)
will not be affected, and will remain in the theory with the same dimension,
as is  required by their conservation equations.

\subsec{Contribution from fields at singularities}

In this case we have two sets of singularities (7-branes)
along two three-spheres in the compact space which
intersect along an $S^1$. 
So, there will be modes coming from the 7-branes and modes coming
from the intersection of the two sets of 7-branes. 
The analysis of the 7-brane modes is similar to what we did in \S4, 
except that now the $S^3$ is really replaced by a singular 
space since one of the angles will have a different periodicity. 
This can be taken into account as we did for the bulk modes in \S3, but
now with $S^3$ replacing $S^5$. Thus, we will again find the same
spectrum described in the previous section arising from each
singularity, but with shifted $U(1)_R$ charges and dimensions.

Next, we turn to the contribution of the intersection region. 
In many cases this region corresponds to a
strongly coupled $d=6$ $\cn=1$ fixed point theory \intersect , about
which not much is known. However, in other cases the low-energy
spectrum is known. 
A particularly simple case is the intersection of
two $D_4$ singularities, which can be described as a $\IZ_2\times
\IZ_2$ orientifold of the type IIB string theory
\refs{\blumzaf,\seninter}. The field theory on the D3-brane in this
case is a $d=4,\cn=1$ theory which, by construction, has an $SL(2,\IZ)$
electric-magnetic duality symmetry. Unfortunately, unlike the case of 
a single $D_4$ singularity
described in the previous section, it is not known how to write down a
Lagrangian for this theory \refs{\seninter,\asty} so we cannot compare
our results with field theory, but our results may help to find a
field theory description for this theory. 
An intersection of this type can occur at any value
of the string coupling, and in particular, at weak coupling one can
perform a perturbative analysis of the spectrum and find that there is
a tensor multiplet living at the intersection. This is a small
multiplet of the $d=6, \cn=1$ SUSY, so again it will give rise to
chiral multiplets in the corresponding $d=4,\cn=1$ SCFT. We will find
one such multiplet for every Kaluza-Klein mode on the $S^1$, since the
tensor multiplet lives on $S^1\times AdS_5$. The $U(1)_R$ symmetry of
the SCFT will include a contribution from the $SO(2)$ symmetry of the
$S^1$, so the corresponding fields will have all possible integer
values of the R charge. The tensor multiplet of $d=6,\cn=1$ includes a
real scalar, an anti-self-dual tensor field and fermions. Upon
reduction on $S^1$ the bosonic spectrum will include the KK modes of a
tensor field (or a vector field which is dual to the tensor in 5
dimensions) and a real scalar field. These fields
naturally fit into a field strength multiplet $W_\alpha$, whose
bosonic modes include a real scalar field $D$ and a tensor field
$F_{\mu \nu}$ (the real scalar field usually appears as an auxiliary
field in gauge theories, but here it is an independent primary
field). We find one such multiplet for every possible value of the
momentum along the $S^1$, up to infinity in the large $N$ limit. The
existence of these multiplets (at least for large $N$) 
is a prediction of the
CFT/AdS correspondence, which may help in finding a Lagrangian
formulation of the CFT in this case.

Similarly, in other intersections of singularities hypermultiplets
arise at the intersection points. These are also small multiplets of
the $d=6, \cn=(1,0)$ SUSY algebra, and in the field theory they would
correspond to the usual chiral primary multiplets, whose bosonic
components are two complex scalar fields $\Phi$ and $F$. In other
cases, the low-energy theory at the intersection of two singularities
can be an interacting conformal theory, whose spectrum is not
well-defined. We will not discuss these cases here.


\newsec{Supergravity Solution for 3-branes in a 7-brane Background}    

A 3-brane probe moving in a 7-brane background
has a low energy effective action which is the same as that
of the exact solution of the corresponding low-energy
field theories \refs{\sw,\swtwo,\senorientifold,\bds}.
In order to argue that the geometry is related to the field theory 
one needs in these cases to resort to a non-renormalization theorem
which explains why the long distance result (gravity solution) can be
continued to short distances (the field theory regime). Generally,
this is valid only when we have at least $\cn=2$ SUSY.
In this section we go beyond the probe approximation and we see
 how the 3-brane deforms the geometry. We will find that if
the number of 3-branes is large the geometry can be
trusted also in the field theory regime.
Therefore with this solution one could 
calculate non-BPS processes involving a wide variety of energy regimes.
We make an ansatz that reduces
 the problem of solving the full supergravity equations to 
solving the  Laplace equation on the background generated by the
7-branes. We were not able to find a full solution of this Laplace
equation, which in principle could be found numerically
 to extract information
about the field theory. We will, however, discuss some qualitative features
of the solution.  The solution has a structure quite similar to 
other cases  where branes can be localized within branes \gaunt . 

We now describe the solution, which boils down to 
a general recipe for introducing a 3-brane in a 7-brane
background.  We first consider a general solution of intersecting
7-branes which preserves some supersymmetry.  
The cases of interest
are (i) A single stack of parallel 7-branes preserving 16 supercharges (or
F-theory on $K3$), 
(ii) Two stacks of parallel 7-branes sharing 5+1 common directions and
preserving a total of 8 supercharges (or F-theory on $CY_3$) and
(iii) Three stacks of 7-branes, with any two stacks sharing 5+1 common
directions and all of them sharing a total of 3+1 common directions,
preserving a total of 4 supercharges (or F-theory on $CY_4$).
Only in the first case (i) the solution is known explicitly. 
In the case (ii) the behaviour of the dilaton is known \seninter\ but 
to our knowledge a general form of the metric is not known explicitly.
In any case we will assume that we have a solution for the
7-branes by themselves. Then we will add the 3-branes. 
In cases (i),(ii) and (iii) the worldvolume field theory 
has $\cn =2, \cn =1$ and $ \cn =1 $
supersymmetry, respectively. In  case (iii) the 3-brane has to 
be chosen with the right orientation so that it does not break
additional supersymmetries. 

Let us assume that we have a supergravity solution for one of the
three cases: it will be of the form
\eqn\sugrasol{ds^{2}= dx_{\parallel}^2 + {g}_{ij}dx^{i}dx^{j},}
where $dx_{\parallel}^2$ is the flat Minkowski 
metric in the directions $0123$.
Furthermore $g$ is a K\"ahler metric and the complexified
 IIB coupling $\tau_{IIB} = \chi + ie^{-\phi}$ is a holomorphic function and
describes, together  with $g_{ij}$, an elliptically fibered CY space.
Since these cases, by assumption,  preserve supersymmetry, there is a spinor
$\epsilon$ which satisfies
\eqn\susyseven{
\eqalign{\delta\lambda &= \Gamma^{M}P_{M}\epsilon^* =0 \cr
\delta\psi_{M} &= (\partial_{M} +
{{1}\over{4}}\omega_{M}^{ab}\Gamma_{ab} - {{i}\over{2}}Q_{M})\epsilon =0,
}}
where we use the notation of \gsw . 
In the cases (i), (ii) and (iii) 
$\tau$ depends on one, two or
three complex variables, respectively. 
We find that the spinors satisfy the
conditions $\Gamma_I \epsilon =0$, where $I$ runs over the complex
variables $z_I$
on which $\tau$ depends, so that we get one, two or three 
conditions on the spinor.
All these conditions are compatible with each other, and they each break
one half of the supersymmetry.


In the $\cn=2$ case (i), the 7-brane metric is explicitly known 
\gsvy 
\eqn\sbmetric{{ g}_{z{\bar z}} = \tau_{2}(z) \mid
\eta^{2}(\tau(z))\prod_{i=1}^{n}(z-z_i)^{-{1\over 12}}dz\mid^2,}
where $z=x^8 + ix^9$, $z_i$ are the positions of the 7-branes, and
$\tau(z)$ is the modular parameter of the elliptic fiber of the 
F-theory compactification. 
  
Now, consider introducing 3-branes into the problem with their worldvolume
spanning the $0-3$ directions.  We make the following  ansatz for the 
Einstein metric\foot{This ansatz  appeared also in \alex\ while this 
paper was in progress.} 
\eqn\metts{
ds^2 = f^{-1/2}dx_{\parallel}^2 + f^{1/2}{g}_{ij}dx^{i}dx^{j}
}
and for the 5-form field
\eqn\ffield{ 
F_{0123i}  =  -{{1}\over{4}}\partial_{i}f^{-1}~. }

The dilaton is the same as in the solution for the 7-branes. 
$f(x^i)$ is a function of the coordinates transverse to the 3-brane.
The self-duality condition  $F = \star_{10} F$ implies that 
 the dual components of
$F_{0123i}$ are  non-zero, and in order to be able to 
solve for a gauge potential we need 
\eqn\laplac{
 {1 \over \sqrt{ g} } 
\partial_{i}(\sqrt{ g}{ g}^{ij}\partial_j f)=
- (2 \pi)^4 N { \delta^6(x-x^0) \over \sqrt{\ g}, }
 }
where we have included a source term at the position of
the $N$ D3-branes so that we produce the right value for the flux
of the five-form field strength. If the 3-branes  are
at different positions we just replace 
$N \delta(x-x^0) \rightarrow \sum_i \delta( x - x^0_i)$.
 
It is easy to see that the ansatz \metts\ \ffield\ leads to a preserved
supersymmetry. We can  write the supersymmetry parameter as 
$\eta = f^{-1/8} \epsilon $, where $\epsilon$ is the supersymmetry preserved
by the 7-branes alone. Then it  is easy to see that all the supersymmetry 
variations, which now include also the 5-form field, vanish provided that 
\eqn\susycond{ (1 + i{\hat\Gamma}_{0123})\eta = 0, }
where $\hat\Gamma$ refers to the flat metric gamma matrices. 
This constraint is 
automatically satisfied in case (iii) with our choice for the charge of
the 3-brane \ffield, while in the other cases half of the SUSYs will
be preserved. Thus, we have verified that our ansatz preserves the
expected amount of supersymmetry, and this, together with \laplac ,
 guarantees that it will also
satisfy the equations of motion.



Now we should discuss more precisely the regime of validity of
the supergravity solution. 
We will be interested here in the field theory limit, 
analogous to the decoupling limit described in detail in \juan .
In that limit  only the local F-theory geometry will be relevant, and it
will encode not only the information about the low-energy effective
action but (according to the conjecture \juan) the full information about
the large $N$ limit of the field theory. To describe the field theory
limit
we will impose the boundary condition that $f \to 0$ when
$x \to \infty $, as opposed to $f \to 1$ which is relevant for the
standard D3+7-branes solution.
 This amounts to taking the  decoupling limit as
in \juan . 
In other words, we are taking the limit $\alpha' \to 0$ keeping the 
mass of the 7-7 , 3-7 and 3-3 strings fixed. 
In order to see when the gravity approximation is valid we first
note that  the equation \laplac\ is linear, so  the solution
for $N$ branes is related to the solution with $N=1$ by $f_N = N f_1$.
When we insert this into our ansatz \metts\ \ffield\ we find, after
defining new variables $x'_{||} = x_{||}/\sqrt{ N }$, that there is
an overall factor of $\sqrt{N}$ in the metric so that by taking
$N$ large the curvature becomes small almost everywhere. The curvature
might blow up at the position of the
7-branes, but this can be taken into account
by adding the fields propagating on the 7-branes (as we did above).

We also need to  understand what happens very close to the point where
the 3-branes are sitting ($x^0$) and what happens at infinity.
If $x^0$ is a non-singular point in the original F-theory geometry, then the 
solution very close to $x^0 $ will behave as $f \sim N/(x-x^0)^4$ and
the geometry locally  will be $AdS_5\times S^5$. This is just saying that we
will have the $\cn =4$ gauge theory in the infrared. 
The behaviour in the ultraviolet (large $x$) 
will depend on the F-theory geometry. 
If the geometry is such that the dilaton asymptotes to a constant 
value then we also have a conformal field theory in the ultraviolet 
which will be like the ones we described in  previous sections.
This physically means that at high energies we
do not distinguish whether the 7-branes and 3-branes are all
together or not. 
Another possibility is that the dilaton goes to zero at infinity. 
It has already been shown in \refs{\bds,\asty,\dls} that the behaviour 
of the dilaton in F-theory is the same as the behaviour of the coupling
constant in the gauge theory. A novel feature of the large $N$ limit
is that the gravity description
is valid in the field theory regime (though it is not {\it perturbative}
field theory). If the dilaton goes to zero at infinity then 
the gravity approximation breaks down as
we go to large distances from the 7-branes. This can be seen by solving 
\laplac\ approximately for large $|z|$, and we find that $f \sim N/(|z|^4 
(\log|z|)^2 )$. Then we see that the square of the Ricci tensor in the
string
frame diverges\foot{It is necessary to go to the string frame to asses the 
validity of the gravity approximations since the dilaton is going to zero. 
Actually the square of the Ricci tensor
goes to a constant in the Einstein frame.}.
%
This  is expected since in these cases at
high energies perturbation theory becomes a good approximation and therefore
gravity should fail. 
The distance at which the curvature diverges grows
exponentially with $N$, mirroring the logarithmic behaviour of the coupling
as a function of energy in the field theory (in a  
 gauge theory of rank $N$
with an $N$-independent one-loop beta function).
At lower energies gravity will still be a good 
approximation if the number of 3-branes is large enough. 
The fact that the geometrical description fails for large distances
for asymptotically free theories in 1+1 and 2+1 dimensions 
was discussed in \juansunny . In the cases discussed in \juansunny\ the
running of the effective coupling was power-like and due to the engineering
dimensions of the coupling. In this four-dimensional case the running
of the coupling is logarithmic so it is more like what is expected
for QCD. We could consider, for example, the theory that arises when we 
start with the $\IZ_2$ orientifold \senorientifold\ 
and we move the four D7-branes away. This corresponds to 
making the fundamental hypermultiplets infinitely massive. We get 
an $\cn =2$, $USp(2N)$ 
gauge theory with a hypermultiplet in the antisymmetric representation. 
The running of the coupling is logarithmic but independent of $N$, since
it is given by the 7-brane solution. This agrees with the field
theory expectation.

We could also have a solution where we have 3-branes sitting on 
a D7-brane. This is a theory which at low energies 
 has the matter content of $\cn = 4$ SYM plus
a fundamental hypermultiplet.  
 In this case the gravity solution will fail in the infrared (close to
the 3-branes)
since the dilaton is going to zero. 
This, of course, is in agreement with the
fact that the field theory becomes free in the infrared in this case.

Note  that the form of the metric \metts\ is such that
the equations for a BPS string, or a string web \refs{\bpsstates,\barton}, 
are the same as if we neglected
the metric generated by the 3-brane (and set $f=1$).
Consider a configuration where the 7-branes are separated from each
other, the 3-branes are  separated from the
7-branes, and  some 3-branes are also separated from
the rest of the 3-branes. In this situation we could have  
 strings (or string  webs)  
going between  different 3-branes, between 3-branes
and 7-branes or between  different 7-branes. 
In the gravity description  the 3-7 strings become strings going  between
the 7-branes and the horizon $U=0$. 
All these types of states are then
states in the field theory, and for large $N$
the proper length of these strings is large so that we can trust
the semiclassical description. This is in contrast with the situation
at small  $N$ where the field theory regime and the semiclassical 
descriptions of the string webs do not overlap. 
Notice that strings going between different 7-branes
can be viewed as the bound 
states of two quarks (hypermultiplets in the fundamental of $USp(2N)$). 
This can be seen starting from a quark-antiquark pair at some distance and
through a description as in \wilson\ one can see that the configuration
would decay into a string going between two different 7-branes. 
The fact that strings going between 7-branes play a role in the field
theory was demonstrated above where we found that they corresponded to 
operators in the 4d CFT (for example, to  the global symmetry currents). 
At the conformal point we do not see the 3-3 strings or the
3-7 strings since
they become strongly interacting massless particles. Of course the 
effect of these interactions is summarized by string theory on the
AdS geometry. 

Another interesting thing to calculate is the K\"ahler metric in the moduli
space of the 3-branes. Consider a 3-brane separated from the
rest of the 3-branes. Its dynamics are described by a Born-Infeld 
action in the supergravity background, which reduces
at low energies to 
\eqn\action{ {\cal L} \sim 
\int \tau_2 F^2 + \tau_1 F\wedge F + g_{ij} \partial X^i \partial X^j. 
}
So, we see that the K\"ahler  metric is just the metric produced by the
7-branes. This gives a method to compute the K\"ahler metric for 
$\cn =1$ theories for $N$ and $g_{YM}^2 N$ large. In principle one
could also compute corrections in $1/\sqrt{g_{YM}^2 N}$ and $1/N$, which are
$\alpha'$ and string loop corrections, respectively. 

\subsec{Three-branes in more general  spaces}

The ansatz that we found above, equations \metts\ and \ffield ,  
is quite general.
Indeed, we can start with {\it any }
 supergravity solution with zero $B$-fields 
(not necessarily supersymmetric) which is the product of four dimensional
Minkowski space and some 
six-dimensional geometry (of the form \sugrasol ),
where the  dilaton field need not be a constant, but all the
fields depend only on the six-dimensional coordinates. 
Then, the solution after we add the 
3-branes has  the form \metts \ffield\ with $f$ given by \laplac .
 Of course, in 
non-supersymmetric cases the solution that we get in this way could
be corrected by $\alpha'$ or loop corrections. These corrections will
be small in the large $N$ limit (if we take $N$ large with everything else
kept fixed). 
 In the decoupling limit, as above, only the local geometry will be 
important and $f \to 0$ when $x \to \infty$. 
These solutions  give the large $N$ limit of field theories that
arise when 3-branes move in the  corresponding geometry.
In fact, we can argue that the solution must have  this form since
we could compactify the three spatial dimensions parallel to the 3-brane
 and
do a U-duality transformation which takes  the 3-brane charge into
momentum. The solution carrying only momentum is 
given in terms of 
    a single harmonic function which satisfies the Laplace 
equation in the transverse space
(in Einstein frame) \hortse\ (supersymmetry was not 
necessary  for the plane wave solutions  studied
in \hortse ).  

In particular, one could study
 3-branes moving on an $ALE$ space or near a singularity in a 
$CY_3$ manifold. 3-branes at various  singularities and orbifold points
 were analyzed at
the conformal point in \refs{\ks,\vafaetal,\oztern,\wittbaryon}.
 The above ansatz provides
a way to solve for the theory away from the conformal point, after
moving the 3-branes away or
blowing up the $A_k$ singularity into a smooth $ALE$ space. In the latter
case there will be strings on the AdS space coming from 3-branes
wrapped on the blown-up 2-cycles, which should have some field
theory interpretation (as found for other branes in \wittbaryon).
Of course it is necessary again to solve the Laplace equation \laplac\
on the corresponding background.
All that we said in the previous section about the validity of the
supergravity solution, etc., goes through also for this case. 
Similarly one could analyze field theories with less supersymmetry.

\subsec{First order approximation}

The equation \laplac\ for the $\cn =2$ case becomes
\eqn\equation{
[g_{z\bar z} \nabla_y^2 + 4  \partial_z \partial_{\bar z}] f = - (2 \pi)^4 
N \delta^2(z-z_0) \delta^4(y),
}
where $g_{z \bar z}$ is given by \sbmetric.
To illustrate an aspect of these solutions
assume that 
$z_0$ is not on any 7-brane. This means
that the metric is regular there and that one can find some new coordinates
$\tilde z$ so that $\tilde z_0 =0$ and such that the metric
has the expansion $
g = 1 + c |\tilde z|^2 + \cdots$. 
Then we can solve the equation \equation\ iteratively by writing $f = 
f_0 + f_1 + \cdots $, where 
\eqn\fzero{
 f_0 = { 4 \pi N \over ({ y}^2 + |\tilde z |^2 )^2, }
}
$f_1$ satisfies
\eqn\iterat{
[\nabla_y^2 +  4  \partial_{ \tilde z} \partial_{\bar {\tilde z}}] 
f_1  = 
 - c |\tilde z |^2 \nabla_y^2 f_0,
}
and so on.
Solving \iterat\ we find that $f_1$ is given by 
\eqn\fone{
f_1 = -{ c \over 6}  { (y^4 + 3 y^2 |\tilde z|^2 + 6 |\tilde z|^4)
 \over (y^2 + |\tilde z|^2) } f_0,
}
so, as expected, when $ \tilde z \to 0$ the correction becomes small.
 From \fone\ one might expect that the leading irrelevant correction is
due to an operator of dimension six. This is not true, however, since
the leading irrelevant correction will come from the fact that that the
dilaton is not constant, and the first derivative of the dilaton at
$z_0$ will give rise to an operator of dimension five 
in the $\bf 6$ representation of
$SO(6)$. This is the leading irrelevant correction to the $\cn=4$
low-energy theory on the 3-branes. In fact, the
coefficient $c$ above is quadratic in the parameter of this perturbation, 
and corresponds to other operators (starting with dimension 6)
which are also induced in this background.

Of course, 
the equation \laplac\ could also be solved numerically. 

\subsec{Solutions in cases of constant coupling}

Consider the $\cn =2$ case when the coupling is constant.
We will see that in this case we can solve \laplac\ (generalizing our
results for the conformal cases above).
The metric transverse to the 7-branes will be of the form 
$ds^2 = g_{\bar z z}  |dz|^2 = \tau_2 |d a|^2 $
where $a$ is the quantity 
appearing in the Seiberg-Witten solution.
Thus, we can define a new variable $w(z)$ such that $dw = \sqrt{\tau_2} da $.
This equation defines a (multivalued) holomorphic function.  
A solution of \laplac\ can then be written as 
\eqn\solhol{ f \sim \sum_i { {N } \over 
( |w - w^{(i)}_0|^2  + y^2 )^2 },
}
where $y$ are the coordinates transverse to the 3-branes and parallel to
the 7-branes  and
$i$ runs over all images of the position of the 3-brane, $z_0$, 
i.e. all the possible values 
$w^{(i)} = w(z_0)$.
In orbifold cases, $ D_4,E_n$, the solutions \solhol\ describe 
branes sitting together in the physical $z$ plane, but of course
separated from the 7-branes. 
In the $H_n$ cases \solhol\ leads to solutions containing several 
groups of branes in the physical $z$ space. These are some solutions 
but not the most general solutions. Solutions that have only one group
of branes in the physical $z$-space can
be gotten by considering functions
of $w$ which, as opposed to \solhol , are not single valued functions of $w$.
These involve the generalized spherical harmonics discussed in section 3.

Similarly we can construct solutions  for other cases of
constant $\tau$ which were described in \barton \foot{
We thank B. Zwiebach for discussions about this.}. As an example consider
the case in which a $D_4$ singularity splits into two $H_1$ singularites
which we set at  positions $z_1 =0$ and $z_2 =1$. In his case
$d a \sim d w \sim  z^{-1/4} (1-z)^{-1/4}dz$ then we see that 
\eqn\solomega{
w \sim  z^{3/4} F(3/4,1/4,7/4,z) 
}
Where $F$ is a hypergeometric function. 
In this case we see that for large $z$ we get $ w \sim z^{1/2}$ which 
is the behaviour at the $D_4$ singularity.
It is simple to take a solution where one has threebranes at different 
points in the physical space by considering eqn. \solhol .
One could find similar solutions in other cases with 
constant dilaton discussed in \barton .

\vskip 1cm

{\centerline {\bf Acknowledgments}}

We would like to thank S. Kachru, D. Morrison, 
A. Strominger, C. Vafa and B. Zwiebach for discussions.
O.A. would like to thank Harvard University and Tel-Aviv University
for their hospitality during the course of this work.

As we were finishing this paper we learned  that T. Hauer was pursuing
ideas similar to those in section 3.

This work was supported in part by 
DOE grant
DE-FG02-96ER40559.

\listrefs{} 

\end 

\newsec{Notes to ourselves}

I just want to write here some calculations, they need not be in the 
paper.

\subsec{Behaviour of $f$ for large $y$ or large $\rho$}

We consider the $\cn = 2 $ case.
We Fourier transform in the coordinates transverse to D3 but parallel 
to 7-branes. We get an equation
\eqn\transf{
 [- k^2 g_{z\bar z} + 4 \partial_z  \partial_{\bar z} ]f_k = \delta^2(z-z^0)
}
For large $y$ we are interested in small $k$ above and we can expand as
explained in some section above and we would see that since $k$ is small
the corrections are small. So if  $y \gg \rho$ we get
\eqn\largey{
f \sim { N \over y^4 }
}
Alternatively for $\rho \gg 1$ we are interested again in small $k$.
When we Fourier transform back we get something of the form
\eqn\dfk{
\int d^4 k e^{ik.y} \sim { 1 \over y} \int_0^\infty dk k^2 J_1(ky)
}
Now for small $k y$ we have $J_1( y k) \sim k y$. 
We are interested in evaluating then
\eqn\fun{
f \sim \int_0^\infty dk k^2 { J_1( ky)
 \over y} f_k(k,\rho)
}
Since we are interested in large $\rho$ we can use a $WKB$ approximation 
to $\transf $ and we get
\eqn\appfk{ f_k  \sim {1 \over \sqrt{\rho} 
( g_{z\bar z})^{1/4} } e^{ - \int^\rho
d\rho' \sqrt{g_{z\bar z}(\rho' )} }
}
This results in  
\eqn\approxf{
f \sim N { 1\over \rho^4 g_{z\bar z}^2}
}
if $g_{z\bar z}$ is slowly varying (slowly than $\rho$) so that we approximate
the integral in \appfk\ by $\rho \sqrt{g_{z\bar z}}$. 
Eventually we are interested
in the case $g_{z\bar z} \sim \log\rho $. 

\subsec{Ricci scalar}

The Ricci scalar is 
\eqn\riccis{
R =  2 f^{-1/2} g^{I \bar J} {\partial_I \tau \partial_{\bar J} \bar \tau 
\over ( \tau_2)^2} \sim { 1 \over (\log \rho)^2  }
}
We see it goes to zero.
However what really determines the validity of the gravity solution
is the curvature in  {\it string } metric. 
Curvatures in string metric are roughly
 $ R_s \sim e^{-\phi/2}( R_E + \nabla^2 \phi) $, but still we see that
it goes to zero.
This does not mean that the metric is non-singular. Notice that already 
in the case of 3-branes the Ricci scalar was zero. 

\subsec{Square of Ricci tensor}

In this section we calculate the square of the Ricci tensor. 
In Einstein metric it is 
\eqn\riccisq{\eqalign{
R_{MN} R^{MN } = &
f^{-1} \left[
2 ( P_i P^*_j g^{ij} )^2 + \right. \cr 
& \kappa^2  \left( 
2  | g^{ij} P_i \partial_j \log f|^2 +  g^{ij} P_i P^*_j g^{kk'}
\partial_k \log f \partial_{k'} \log f  \right) \cr 
& \left.  \kappa^4  6  \left( g^{ij } \partial_i \log f \partial_j 
\log f \right)^2  \right]
}}
Now we see that the last term goes like a constant and the other terms are
decrease for large $\rho$. Notice that this last term is the  same as 
what we would have without 7-branes. 
But now if we calculate this in string units we would find that it becomes
large since $R^2_s \sim e^{ - \phi } ( R_E^2  +  \cdots ) $ and
$e^{-\phi} \sim \log \rho $.

Notice an interesting feature of the expression \riccisq\ which is that
there is an overall factor of $f^{-1}$, so when we scale up $N$ keeping 
everything else fixed we see that the curvature becomes small everywhere.

So in conclusion we see that the curvatures in string units blow up
as the dilaton becomes small.

\subsec{First order approximation}

Now we consider a region of the space where the coupling is running
logarithmically $ \tau \sim  { i n \over 2 \pi}  \log|z| $. Then, consider
3-branes sitting at some position $z_0\gg1$, so that we are in the 
asymptotic region. The equation for the function $f$ reduces to
\eqn\equation{
[g_{z\bar z} \nabla_y^2 + 4  \partial_z \partial_{\bar z}] f = (2 \pi)^4 
N \delta^2(z-z_0) \delta^4(y)
}
Then we can solve the equation iteratively by writing $g_{z\bar z} =
g_{z\bar z} (z_0) + h(z-z_0,\bar z -\bar z_0) $ and treating $h$ as
a small perturbation and 
defining $f = \sum_n f_n$ where
\eqn\iterat{
[\nabla_{\tilde y}^2 +  4  \partial_z \partial_{\bar z}] f_n  = 
 - h(z-z_0,\bar z -\bar z_0) \nabla_y^2 f_{n-1}~~~~~~~~{ n\ge 1}
}
where  ${\tilde y} \sim y^2/\sqrt{g_{z \bar z}(z_0)}$ and the zeroth
order solution is 
\eqn\fzero{
 f_0 = { 4 \pi N \over ({\tilde y}^2 + |z-z_0|^2 )^2 }
}
Taking $g_{z \bar z} \sim \log|z|$ we can find the first order solution
by expanding the logarithm in powers and then resuming the expansion.
We get
\eqn\fone{
f_1 = {const(z_0)} (- 16) 
\Re\left[ ( 1 + 1/\tilde z ) \log(1 + \tilde z) -1 \right]
{ | z-z_0 |^2 \over (\tilde y^2 + |z-z_0|^2 )^3}
}
where $\tilde z = (z-z_0)/z_0$. 
We see that for small $\tilde z$ this first order correction
goes like $ 1/\tilde z^3 $ (for $y =0$), so that it is subleading
compared with \fzero . This is precisely what we expect, 
\fone\ takes into account the leading irrelevant corrections 
of the IR $\cn =4$ superconformal field theory. 
We can also trust $f_1$ if we want to calculate the leading corrections
in the region $y\gg z$, where again it gives a subleading correction
compared to \fzero .
In principle we could continue expanding and we would get 
further corrections.

The equation \laplac\ could also be solved numerically in principle.